\documentclass[11pt]{article}

\usepackage{amsfonts}
\usepackage{amsmath}
\usepackage{amssymb}
\usepackage{graphicx}

\numberwithin{equation}{section}

\newcommand{\der}{\mathrm{d}}

\newtheorem{theorem}{Theorem}[section]

\newtheorem{example}[theorem]{Example}

\newtheorem{remark}[theorem]{Remark}

\title{Phase Splitting for Periodic Lie Systems}
\author{R. Flores-Espinoza$^{\ast}$, J. de Lucas$^{\dagger}$ and Yu. M. Vorobiev$^{\ast}$}

\date{$^{\ast}$Departamento de Matem\'aticas, Universidad de Sonora, M\'exico\\
$^{\dagger}$Departamento de F\'{\i}sica Te\'orica, Universidad de Zaragoza, Spain\\ and\\
Institute of Mathematics of the Polish Academy of Sciences, Warszawa, Poland}

\begin{document}

\maketitle

\begin{abstract}
In the context of the Floquet theory, using a variation of parameter
argument, we show that the logarithm of the monodromy of a real periodic Lie
system with appropriate properties admits a splitting into two parts, called
dynamic and geometric phases. The dynamic phase is intrinsic and linked to
the Hamiltonian of a periodic linear Euler system on the co-algebra. The
geometric phase is represented as a surface integral of the symplectic form
of a co-adjoint orbit.

\end{abstract}

\section{Introduction}

The so-called Lie systems, representing a special sort of (generally nonlinear) time-dependent dynamics on
$G$-spaces, are of a great interest in the integrability theory of non-autonomous systems of ODEs as well as in
various physical applications (see, for example, \cite{CGM1,CGM2, CLR, CaLu10, CaLu09} and references therein). In
this paper, we are interested in periodic Lie systems in the context of the Floquet theory and
the phase splitting problem. The original motivation for this problem came from quantum mechanics \cite{Be84} and then, for classical integrable systems, the phase phenomena was studied in \cite{Han}. A general geometric approach for computing the geometric and dynamic
phases for Hamiltonian systems with symmetries was developed in \cite{MMR}. Our goal is to give a
version of phase splitting for periodic Lie systems which is based on a variation of parameter
argument \cite{FV, Vor}. First, we observe that the geometric and dynamic phases (as elements of a
Lie algebra) are naturally defined for every smooth one-parameter family of uniformly reducible
periodic Lie systems containing the trivial system. Then, we show that an individual periodic
Lie system which is reducible and contractible, is included in such a family and hence, the
logarithm of its monodromy is the sum of two parts, called the dynamic and geometric phases. The
dynamical phase is defined in an intrinsic way and linked to the time-dependent Hamiltonian of a
periodic linear Euler system on the co-algebra. The geometric phase is given as a surface integral
of the symplectic form of a co-adjoint orbit. This result can be viewed as a generalization of
``scalar'' formulas for the dynamic and geometric phases of cyclic solutions to periodic linear
Euler systems \cite{FV, Vor}. Moreover, in the case when a Lie system comes from a simple
mechanical system, we show that our result agrees with known phase formulas in the reconstruction
theory for Hamiltonian systems with symmetries \cite{Bla, MMR}.

The question on the phase splitting naturally appears in the spectral problems for quantum systems in the semiclassical approximation \cite{KarVor, LJ}. Here, the classical geometric phase defines a correction to the Bohr-So\-mmer\-feld quantization rule and the dynamical phase gives an excitation of the classical energy level. In this context, one of the possible applications of our results is related to the computation of semiclassical spectra of spin-like quantum systems whose classical limits are just periodic Lie systems.

\section{The Floquet theory for periodic Lie systems}

The classical Floquet theory for linear periodic systems \cite{YaSt} can be naturally extended to
the nonlinear case. Suppose we start with a time-periodic dynamical system on a smooth manifold $M$:
\begin{equation}
\frac{\der x}{\der t} = X_{t}(x),   \qquad x\in M. \label{E1}%
\end{equation}
Here $X_{t}$ is a smooth time-dependent vector field on $M$ which is $2\pi$-periodic in $t$,
$X_{t+2\pi}(x) = X_{t}(x)$. We assume that $X_{t}$ is complete. Let $\mathrm{Fl}^{t}: M \rightarrow M$
be the flow of $X_{t}$,
\[
\frac{\der \mathrm{Fl}^{t}(x)}{\der t} = X_{t} \circ \mathrm{Fl}^{t}(x), \qquad
\mathrm{Fl}^{0} = \mathrm{id}_{M}.
\]
From the periodicity condition and standard arguments, it follows that
\begin{equation} \label{M1}
\mathrm{Fl}^{t + 2\pi} = \mathrm{Fl}^{t} \circ \mathrm{Fl}^{2 \pi} \qquad \forall \, t \in\mathbb{R}.
\end{equation}
The diffeomorphism $\mathrm{Fl}^{2 \pi} \in \mathrm{Diff}(M)$ is called the \textit{monodromy} of
(\ref{E1}). System (\ref{E1}) is said to be \textit{reducible } if there exists a time-dependent
diffeomorphism on $M$, $2\pi$-periodic in $t$, which transforms the system into an autonomous one.
Observe that the reducibility property is equivalent to the \textit{Floquet representation} for
the flow of $X_{t}$:
\begin{equation}
\mathrm{Fl}^{t} = P^{t} \circ \Xi^{t}, \label{E2}
\end{equation}
where $\Xi^{t} : M \rightarrow M$ is a one-parameter group of diffeomorphisms,
$\Xi^{t + \tau} = \Xi^{t} \circ \Xi^{\tau}$ and $P^{t} : M \rightarrow M$ is a time-dependent
diffeomorphism such that $P^{t + 2\pi} = P^{t}$. Indeed, if there exists a $2\pi$-periodic change
of variables
\begin{equation}
x \mapsto y = (P^{t})^{-1}(x), \qquad P^{0} = \mathrm{id}_{M}, \label{T1}
\end{equation}
transforming (\ref{E1}) to the system
\begin{equation}
\frac{\der y}{\der t} = Y(y),      \label{E3}
\end{equation}
with time-independent vector field
\begin{equation}
Y = \frac{\der (P^{t})^{-1}}{\der t} \circ P^{t} + (P^{t})^{-1}_* X_{t},          \label{E4}
\end{equation}
then (\ref{E2}) holds, where the one-parameter group $\Xi^{t}$ is defined as the flow of $Y$.
Conversely, if the flow admits decomposition (\ref{E2}) for some $\Xi^{t}$ and $P^{t}$, then the
$2\pi$-periodic transformation (\ref{T1}) reduces system (\ref{E1}) to the form (\ref{E3}) with
$\displaystyle Y(y) = \frac{\der}{dt} \Big|_{t=0} \, \Xi^{t}(y)$. We get the following reducibility
criterion: time-periodic system (\ref{E1}) is reducible if and only if there exists a one-parameter
group of diffeomorphisms $\Xi^{t} : M \rightarrow M$ such that
\begin{equation}
\mathrm{Fl}^{2\pi} = \Xi^{2\pi}.                 \label{R1}
\end{equation}
Therefore, as in the linear case, the reducibility of system (\ref{E1}) is completely controlled by
its monodromy.

It is also useful to introduce the notion of relative reducibility. Let
$\mathcal{G} \subset \mathrm{Diff}(M)$ be a subgroup of diffeomorphisms on $M$. Assume that the
flow $\mathrm{Fl}^{t}$ takes values in $\mathcal{G}$, that is, $\mathrm{Fl}^{t} \in \mathcal{G}$
for all $t \in \mathbb{R}$. Then, we say that system (\ref{E1}) is \textit{reducible relative to}
$\mathcal{G}$ (or, shortly, $\mathcal{G}$-\textit{reducible}), if one can choose a reducibility,
time-dependent, diffeomorphism $P^{t}$ in (\ref{T1}), (\ref{E3}) to be a $2\pi$-periodic curve in
the subgroup $\mathcal{G}$. The criterion above is modified as follows: the $\mathcal{G}$-reducibility
of system (\ref{E1}) is equivalent to the existence of a one-parameter subgroup $\{\Xi^{t}\}$ in
$\mathcal{G}$ which passes through the monodromy at $t = 2\pi$.

In general, the question on the embedding of the monodromy into a one-parameter group of diffeomorphisms
is a difficult task. Our point is to discuss this problem for a special class of time dependent dynamical
systems on $G$-spaces, namely, the \textit{Lie systems} \cite{CGM1}.

Suppose that the manifold $M$ is endowed with a smooth left action $\Phi:G\times M\rightarrow G$ of
a real connected Lie group $G$. For every $g\in G$, denote by $\Phi_{g} : M \rightarrow M$ the
diffeomorphism given by $\Phi_{g}(x) = \Phi(g,x)$ for all $x\in M$. Fixing $x \in M$, we define also
the smooth mapping $\Phi^{x} : G \rightarrow M$ letting $\Phi^{x}(g) = \Phi(g,x)$. Let $\mathfrak{g}$
be the Lie algebra of $G$. By a \textit{periodic Lie system} on $M$ associated with the $G$-action
we mean the following non-autonomous system
\begin{equation}
\frac{\der x}{\der t} = T_{e} \Phi^{x}(\phi(t)),  \qquad x\in M \label{Li1}
\end{equation}
where $T_{e}\Phi^{x} : \mathfrak{g} \rightarrow T_{x}M$ is the tangent map of $\Phi^{x}$ at the identity
element $e$ and $\mathbb{R} \ni t \mapsto \phi(t) \in\mathfrak{g}$ is a smooth $2\pi$-periodic curve in
the Lie algebra, i.e. $\phi(t + 2\pi) = \phi(t)$. The vector field $X_{t}(x) = T_{e} \Phi^{x}(\phi(t))$
of this system is represented as a linear combination of the infinitesimal generators of the $G$-action with
time-periodic coefficients. In general, $X_{t}$ is not $G$-invariant but the trajectories of (\ref{Li1})
belong to the orbits of the $G$-action. Let $L_{g} : G \rightarrow G$ and $R_{g} : G\rightarrow G$
denote the left and right translations by an element $g\in G$, respectively. One can associate to (\ref{Li1})
the following non-autonomous system in $G$
\begin{equation}
\frac{\der g}{\der t} = T_{e} R_{g}(\phi(t)), \qquad g\in G.       \label{Li2}
\end{equation}
System (\ref{Li2}) is complete because its vector field is time-periodic and right invariant. Since the
right invariant vector fields are the infinitesimal generators of the action of $G$ on itself by the left
translations, system (\ref{Li2}) can also be viewed as a periodic Lie system associated with this
left $G$-action. Consider the solution $\mathbb{R} \ni t \mapsto f(t) \in G$ to the initial value problem
\begin{equation}
\frac{\der f}{\der t} = T_{e} R_{f}(\phi(t)), \qquad f(0)=e.      \label{Fan1}%
\end{equation}
Then, we have the following relationship between the flow $\mathrm{Fl}^{t}$ of system (\ref{Li1}) and
the solution $f(t)$ of system (\ref{Fan1}):
\begin{equation}
\mathrm{Fl}^{t} = \Phi_{f(t)}.              \label{Fl1}%
\end{equation}
In particular, the flow of system (\ref{Li2}) is given by $g \mapsto L_{f(t)} g$. In analogy with the
linear case, we will call $f(t)$ the \textit{fundamental solution} of the periodic Lie system (\ref{Li2})
on $G$. In terms of the fundamental solution, the property (\ref{M1}) for the flow of (\ref{Li2})
reads $f(t + 2\pi) = f(t)\cdot m$. Here $m := f(2\pi) \in G$ is said to be the
\textit{monodromy element} of (\ref{Li2}). It follows from (\ref{Fl1}) that the monodromy of system
(\ref{Li1}) is represented as $\mathrm{Fl}^{2\pi} = \Phi_{m}$. Let
$\mathcal{G}_{G,\Phi} =\{\Phi_{g}$, $g\in G \}$ be the subgroup of diffeomorphisms on $M$ generated
by the $G$-action. Then, formula (\ref{Fl1}) says that $\mathrm{Fl}^{t} \in \mathcal{G}_{G,\Phi}$ and
hence one can talk on the reducibility of (\ref{Li1}) relative to the group $\mathcal{G}_{G,\Phi}$.

Assume that the monodromy element $m$ lies in the image of the exponential map
$\exp : \mathfrak{g} \rightarrow G$,%
\begin{equation}
m = \exp k,                         \label{Re1}%
\end{equation}
for a certain $k \in \mathfrak{g}$. This implies that the Floquet representation for the fundamental solution reads
\begin{equation}
f(t) = L_{p(t)}\left( \exp\left( \frac{tk}{2\pi} \right) \right),       \label{Floq}%
\end{equation}
where $t \mapsto p(t)$ is a $2\pi$-periodic curve in $G$ with $p(0) = e$. It follows from here and
(\ref{Fl1}) that the monodromy $\mathrm{Fl}^{2\pi}$ satisfies condition (\ref{E2}) for the one-parameter
group of diffeomorphisms $\Xi^{t} = \Phi_{\exp(\frac{tk}{2\pi})}$ and hence system (\ref{Li1}) is
$\mathcal{G}_{G,\Phi}$-reducible. Therefore, under the $2\pi$-periodic change of variables (\ref{T1})
with
\[
P^{t} = \mathrm{Fl}^{t} \circ \Phi_{\exp(-\frac{tk}{2\pi})} = \Phi_{p(t)},%
\]
system (\ref{Li1}) is transformed to the autonomous Lie system of the form
${\der y}/{dt} = T_{e} \Phi^{y}(\frac{k}{2\pi})$. Condition (\ref{Re1}) becomes also necessary
for the reducibility under a natural assumption on the $G$-action. We have the following
criterion: if the action $\Phi$ of $G$ on $M$ is \textit{faithful}, then property (\ref{Re1}) is a
sufficient and necessary condition for the $\mathcal{G}_{G,\Phi}$-reducibility of system (\ref{Li1}).
In particular, periodic Lie system (\ref{Li2}) is reducible relative to the group of left translations on
$G$ if and only if (\ref{Re1}) holds.

One can also show that if the $G$-action is not faithful, then the criterion of the
$\mathcal{G}_{G,\Phi}$-reducibility for system (\ref{Li1}) leads to the following representation for
the monodromy element:%
\begin{equation}
m = m_{0} \cdot\exp k,          \label{Re2}%
\end{equation}
for a certain element $m_{0}$ in the kernel of the homomorphism $g\mapsto \Phi_{g}$. In this case,
system (\ref{Li2}) in $G$ is not necessarily reducible. In terms of the monodromy of (\ref{Li1}),
reducibility condition (\ref{Re2}) reads $\mathrm{Fl}^{2\pi} = \Phi_{\exp k}$.

Consider the following important class of periodic Lie systems associated with linear
representations of $G$. Let $\mathrm{Ad} : G \times \mathfrak{g} \rightarrow \mathfrak{g}$ be the
adjoint action  of the Lie group $G$ on its Lie algebra,
$\mathrm{Ad}_{g} = T_{g}R_{g^{-1}} \circ T_{e}L_{g}$. Taking
$\Phi = \mathrm{Ad}$ and $M = \mathfrak{g}$ for (\ref{Li1}), we get the following Lie system
\begin{equation}
\frac{\der x}{\der t} = \mathrm{ad}_{\phi(t)}x, \qquad x \in \mathfrak{g},    \label{Eu1}%
\end{equation}
which is called a \textit{periodic linear Euler system} on $\mathfrak{g}$. Here
$\mathrm{ad}_{\phi} : \mathfrak{g} \rightarrow \mathfrak{g}$ is the adjoint operator,
$\mathrm{ad}_{\phi} y = [\phi,y]$. It follows from (\ref{Fl1}) that the flow of (\ref{Eu1}) is
$\mathrm{Fl}^{t} = \mathrm{Ad}_{f(t)}$, where $f(t)$ is the fundamental solution in (\ref{Fan1}).
Therefore, $\operatorname*{Fl}{}^{t}$ takes values in the adjoint group $\mathrm{Ad}\,\mathfrak{g}$ of
the Lie algebra $\mathfrak{g}$, which is generated by the elements $\exp(\mathrm{ad}_{z})$, for
$z \in \mathfrak{g}$. Since $G$ is connected, the kernel of the adjoint representation
$g \mapsto \mathrm{Ad}_{g}$ coincides with the center $Z(G)$ of the Lie group. Then, the
$\mathrm{Ad}\, \mathfrak{g}$-reducibility of linear Euler equation (\ref{Eu1}) is equivalent to
representation (\ref{Re2}) for some $m_{0} \in Z(G)$ and $k \in \mathfrak{g}$. In terms of the
monodromy $\mathfrak{M} = \mathrm{Ad}_{m}$ of (\ref{Eu1}), the reducibility criterion says that
$\mathfrak{M} = \exp(\mathrm{ad}_{k})$.

Remark that the reducibility condition (\ref{Re1}) automatically holds in the case when $G$ belongs
to the class of \textit{exponential} Lie groups which includes, for example, the Lie groups of
\textit{compact type} \cite{DH, DK, MS}. If the Lie group $G$ is not exponential, then one can
apply the following criterion. Assume that the monodromy element $m$ is \textit{regular} and the
isotropy subgroup $G_{m} = \{\alpha \in G \; \big| \; \alpha \cdot m = m \cdot\alpha \}$ is
\textit{connected}. Then, it follows that $G_{m}$ is \textit{abelian} \cite{DK} and hence,
(\ref{Re1}) holds.

\begin{example}
Let $G = {SO}(3)$ be the compact Lie group of all orthogonal $3\times3$ matrices $g$ with $\det g = 1$
and $\mathfrak{g} = \mathfrak{so}(3)$ its Lie algebra of skew-symmetric matrices. A periodic Lie system
on ${SO}(3)$ is written as
\begin{equation}
\frac{\der g}{\der t} = (\Lambda\circ w(t))\cdot g,     \label{Rot1}%
\end{equation}
where $t \mapsto w(t) \in \mathbb{R}^{3}$ is a $2\pi$-periodic vector function and $\Lambda\circ x$
denotes the matrix of the cross product in $\mathbb{R}^{3}$, $(\Lambda \circ x)y = x\times y$. Under
the identification of $\mathfrak{so}(3)$ with $\mathbb{R}^{3}$, the corresponding periodic linear Euler
system takes the form
\begin{equation}
\frac{\der x}{\der t} = w(t) \times x, \qquad x \in \mathbb{R}^{3}.   \label{Rot2}%
\end{equation}
Since the monodromy element $m \in {SO}(3)$ of {\upshape (\ref{Rot1})} is a rotation in $\mathbb{R}^{3}$,
we  have $m = \exp \Lambda \circ \nu$ for $\nu \in \mathbb{R}^{3}$. Therefore, systems
{\upshape (\ref{Rot1})} and {\upshape (\ref{Rot2})} are reducible.
\end{example}

The next example is related to the reducibility of periodic linear Hamiltonian systems on $\mathbb{R}^{2}$
\cite{YaSt}.

\begin{example}
Consider the special linear group $G = {SL}(2;\mathbb{R})$ of all real $2 \times 2$ matrices with
determinant one. The corresponding Lie algebra $\mathfrak{g} = \mathfrak{sl}(2;\mathbb{R})$
consists of traceless $2 \times 2$ matrices. A periodic Lie system in ${SL}(2;\mathbb{R})$ is of the
form%
\begin{equation}
\frac{\der g}{\der t} = \left[
\begin{array}{cc}%
a_{1}(t) & a_{2}(t)\\
a_{3}(t) & -a_{1}(t)
\end{array}
\right]  \cdot g,           \label{Sp1}%
\end{equation}
where $a_{i}(t)$ $(i=1,2,3)$ are $2\pi$-periodic, real functions. It is well-known {\upshape \cite{DH, DK}}
that the exponential  map for ${SL}(2;\mathbb{R})$ is not surjective. The monodromy element $m \in
{SL}(2;\mathbb{R})$ of {\upshape (\ref{Sp1})} has the representation $m = \pm \exp k$, for $k =
\left[
\begin{array}{cc}%
k_{1} & k_{2}\\
k_{3} & -k_{1}%
\end{array}
\right]$. Moreover, $m$ is in the image of the exponential map if and only if $ \mathrm{tr}\; m  > -2$ or
$m = -I$. In the opposite case, system {\upshape (\ref{Sp1})} is not reducible. Identifying
$\mathfrak{sl}(2, \mathbb{R})$ with $\mathbb{R}^{3}$, we can write the periodic linear Euler system
associated with {\upshape (\ref{Sp1})} in the form
\begin{equation}
\frac{\der x}{\der t} = \mathcal{I}(w(t) \times x),    \qquad x\in \mathbb{R}^{3},   \label{Sp2}%
\end{equation}
where $\mathcal{I} = \mathrm{diag}(1,1,-1)$ and $w(t)=(2a_{1}(t),-a_{2}(t)-a_{3}(t),a_{2}(t)-a_{3}(t))$.
The kernel of the adjoint representation of ${SL}(2;\mathbb{R})$ is the two element group $\{I,-I\}$.
The adjoint group of $\mathfrak{sl}(2;\mathbb{R})$ is isomorphic to the Lorentz group ${SO}^{+}(2,1)$
which is exponential {\upshape \cite{DH}}. Therefore, linear Euler system {\upshape (\ref{Sp2})} is
reducible since for its  monodromy $\mathfrak{M} \in {SO}^{+}(2,1)$ we have
$\mathfrak{M} = \exp \mathcal{I}(\Lambda\circ v)$,  for
$v=(2k_{1},-k_{2} - k_{3},k_{2} - k_{3}) \in \mathbb{R}^{3}$.
\end{example}

\section{The Mapping D}

Denote by $C_{e}^{\infty}(\mathbb{R},G)$ the set of all smooth curves
$\alpha(t) : \mathbb{R} \rightarrow G$ in the Lie group with $\alpha(0)=e$, and
$C^{\infty}(\mathbb{R},\mathfrak{g)}$ the set of all smooth curves in the Lie algebra $\mathfrak{g}$.
Introduce the mapping $\mathrm{D} : C_{e}^{\infty}(\mathbb{R},G) \rightarrow C^{\infty} (\mathbb{R},\mathfrak{g)}$
given by
\[
\mathrm{D} \alpha(t) := T_{\alpha(t)} R_{\alpha(t)^{-1}} \left( \frac{\der \alpha}{\der t}(t) \right)
\in \mathfrak{g}.
\]
Then, in terms of $\mathrm{D}$, equation (\ref{Fan1}) for the fundamental solution $f(t)$ is rewritten
as follows
\begin{equation}
\mathrm{D} f = \phi.        \label{DF1}%
\end{equation}
Moreover, for any $a \in\mathfrak{g}$ and $\alpha, \beta \in
C_{e}^{\infty} (\mathbb{R},G)$, the following identities hold \cite{DK}:
\begin{eqnarray}
\mathrm{D}(\exp ta) & = & a,        \label{Pr1} \\
\mathrm{D} \alpha^{-1}(t) & = & - \mathrm{Ad}_{\alpha^{-1}(t)} \mathrm{D} \alpha(t), \label{Pr2}\\
\mathrm{D}(L_{\alpha} \beta)(t) & = & \mathrm{D} \alpha(t) + \mathrm{Ad}_{\alpha(t)}
\mathrm{D}\beta(t),     \label{Pr3}  \\
\frac{\der }{\der t} \mathrm{Ad}_{\alpha(t)} & = & \mathrm{ad}_{\mathrm{D} \alpha(t)} \circ
\mathrm{Ad}_{\alpha(t)}. \label{Pr4}%
\end{eqnarray}

Let $\sigma$ be a parametrized surface in $G$ given by a smooth map
$\mathbb{R}^{2}\ni(s,t) \mapsto \sigma(s,t)\in G$. Denote by $\mathrm{D}_{s}$ and $\mathrm{D}_{t}$ the
mappings which act on the $s$-parameter and $t$-parameter families of curves $\sigma_{t}$ and $\sigma_{s}$
associated with $\sigma$, respectively,%
\[
\mathrm{D}_{s} \sigma(s,t) := \mathrm{D} \sigma_{t}(s) \equiv
T_{\sigma(s,t)}R_{\sigma(s,t)^{-1}} \left( \frac{\partial\sigma(s,t)}{\partial s} \right),
\]%
\[
\mathrm{D}_{t} \sigma(s,t) := \mathrm{D} \sigma_{s}(t) \equiv
T_{\sigma(s,t)}R_{\sigma(s,t)^{-1}} \left( \frac{\partial\sigma(s,t)}{\partial t} \right).
\]
One can show that the relationship between these two mappings is given by the
``zero curvature'' type equation:%
\begin{equation}
\frac{\partial\mathrm{D}_{t} \sigma}{\partial s} - \frac{\partial
\mathrm{D}_{s}\sigma}{\partial t} + [\mathrm{D}_{t} \sigma,\mathrm{D}_{s} \sigma] = 0. \label{Cur}%
\end{equation}

\section{Dynamic and Geometric Phases}  \label{sec:dyngeom}

Suppose we start with a family of periodic Lie systems in $G$ of the form (\ref{Li2}) associated
with a $s$-parameter family $\{\phi_{s}\}$ of closed curves in $\mathfrak{g}$ given by a
$C^{\infty}$ mapping $[0,1] \times \mathbb{R} \ni(s,t) \mapsto \phi(s,t) \in\mathfrak{g}$ with
\begin{eqnarray}
\phi(s,t+2\pi) & = & \phi(s,t), \label{Hom1} \\
\phi(0,t) & = & 0. \label{Hom2}%
\end{eqnarray}
Let $f(s,t)$ be the parameter dependent fundamental solution,%
\begin{equation}
\frac{\der f(s,t)}{\der t} = T_{e} R_{f(s,t)}(\phi(s,t)), \qquad f(s,0) = e.  \label{FP1}%
\end{equation}
It is clear that $f(s,t)$ is smooth in both variables $s$ and $t$. Moreover, $f(0,t) = e$ because
of (\ref{Hom2}). Assume that the family of periodic Lie systems is uniformly reducible, that is,
for every $s \in [0,1]$, the monodromy element has the representation
\begin{equation}
m(s) = f(s,2\pi) = \exp k(s),          \label{MP1}%
\end{equation}
for a certain $k(s) \in \mathfrak{g}$, smoothly varying in $s$ and such that $k(0)=0$. Consider
the $G$-valued function
\begin{equation}
p(s,t) = L_{f(s,t)} \exp \left( -\frac{tk(s)}{2\pi} \right),   \label{Pe1}%
\end{equation}
with properties
\begin{eqnarray}
p(s,t + 2\pi) & = & p(s,t),      \label{Pe2}\\
p(s,0) & = & e,             \label{Pe3}\\
p(0,t) & = & e. \label{Pe4}
\end{eqnarray}
Applying the mapping $\mathrm{D}_{t}$ to both sides of (\ref{Pe1}) and by using
(\ref{DF1}), (\ref{Pr1}) and (\ref{Pr3}), we derive the identity%
\[
\frac{k(s)}{2\pi} = \mathrm{Ad}_{p(s,t)^{-1}} \bigl( \phi(s,t) - \mathrm{D}_{t}p(s,t) \bigr).
\]
Integrating this equality in $t$ over $[0,2\pi]$ gives
\begin{equation}
k(s) = \int_{0}^{2\pi} \mathrm{Ad}_{p(s,t)^{-1}} \bigl(\phi(s,t) \bigr) \der t - \int_{0}^{2\pi}
\mathrm{Ad}_{p(s,t)^{-1}} \bigl( \mathrm{D}_{t}p(s,t) \bigr) \der t. \label{For}%
\end{equation}
By using (\ref{Pr2}), (\ref{Cur}) and (\ref{Pe4}), we compute the second term in (\ref{For}):
\begin{align*}
\int_{0}^{2\pi} \mathrm{Ad}_{p(s,t)^{-1}} & \bigl( \mathrm{D}_{t}p(s,t) \bigr) \der {t}
 = -\int_{0}^{2\pi} \mathrm{D}_{t}p^{-1}(s,t) \der {t}\\
& = - \int_{0}^{2\pi} \int_{0}^{s} \frac{\partial}{\partial u}(\mathrm{D}_{t} p^{-1}(u,t)) \der {u} \der {t}
+ \int_{0}^{2\pi} \mathrm{D}_{t}p^{-1}(0,t) \der {t}\\
& = \int_{0}^{s} (- \mathrm{D}_{u}p^{-1}(u,2\pi) + \mathrm{D}_{u}p^{-1}(u,0)) \der {u}\\
& \qquad - \int_{0}^{2\pi} \int_{0}^{s} [\mathrm{D}_{u}p^{-1}(u,t), \mathrm{D}_{t}p^{-1}(u,t) ] \der {u} \der {t}.
\end{align*}
The first summand in the last formula vanishes because of (\ref{Pe2}), (\ref{Pe3}). Summarizing, we get
the following result.

\begin{theorem}
For every $s \in [0,1]$, the $\log$ phase $k(s)$ in {\upshape (\ref{MP1})} has the decomposition
\begin{equation}
k(s) = k_{\mathrm{dyn}}(s) + k_{\mathrm{geom}}(s), \label{Lo1}%
\end{equation}
where%
\begin{align}
k_{\mathrm{dyn}}(s) & =  \int_{0}^{2\pi} \mathrm{Ad}_{p^{-1}(s,t)} \phi(s,t) \der {t}, \label{Lo2}\\
  &   \nonumber\\
k_{\mathrm{geom}}(s) & = \int_{0}^{2\pi} \int_{0}^{s} [\mathrm{D}_{u}p^{-1}(u,t),\mathrm{D}_{t}p^{-1}(u,t)]
\der {t}\der {u}. \label{Lo3}%
\end{align}
\end{theorem}

The components $k_{\mathrm{dyn}}(s)$ and $k_{\mathrm{geom}}(s)$ will be called the \textit{dynamic}
and \textit{geometric} \textit{phases} of the family of periodic Lie systems, respectively. Now,
let us give interpretations of $k_{\mathrm{dyn}}(s)$ and $k_{\mathrm{geom}}(s)$ in terms of the
Poisson geometry and Hamiltonian dynamics on the dual space (the co-algebra) $\mathfrak{g}^{\ast}$
of $\mathfrak{g}$. Let $\Phi : G \times \mathfrak{g}^{\ast} \rightarrow \mathfrak{g}^{\ast}$ be the
left action of $G$ on the co-algebra $\mathfrak{g}^{\ast}$ given by
$\Phi_{g} = \mathrm{Ad}_{g^{-1}}^{\ast}$, where
$\mathrm{Ad}_{\alpha}^{\ast} : \mathfrak{g}^{\ast} \rightarrow \mathfrak{g}^{\ast}$ is the co-adjoint
action of the Lie group. Then, Lie system (\ref{Li1}) associated to such action and the function
$\phi = \phi(s,t)$ gives the family of periodic linear Euler systems on $\mathfrak{g}^{\ast}$:
\begin{equation}
\frac{\der \xi}{\der {t}} = - \mathrm{ad}_{\phi(s,t)}^{\ast}\xi, \qquad \xi \in
\mathfrak{g}^{\ast}.                                                        \label{CoE}%
\end{equation}
It follows from (\ref{Fl1}) that the flow (the fundamental solution) of (\ref{CoE}) is given by
$\mathrm{Fl}^{t} = \mathrm{Ad}_{f^{-1}(s,t)}^{\ast}$, where $f(s,t)$ is the $G$-valued fundamental
solution in (\ref{Fan1}). Taking into account (\ref{Fl1}), we get that the monodromy of (\ref{CoE})
is of the form $\mathrm{Fl}^{2\pi} = \exp (-\mathrm{ad}_{k(s)}^{\ast})$. For every $s$, system
(\ref{CoE}) represents a time-dependent Hamiltonian system relative to the ``plus'' Lie-Poisson
bracket on $\mathfrak{g}^{\ast}$ \cite{MR}  and the function $H_{t}(\xi) = - \langle \xi,\phi(s,t)
\rangle$. Here, we denote by $\langle \, {,} \, \rangle$ the pairing between vectors and covectors.
Pick a point $\mu \in\mathfrak{g}^{\ast}$. Then, in terms of the time-dependent Hamiltonian $H_{t}
: \mathfrak{g}^{\ast} \rightarrow \mathbb{R}$, we have the following representation for the
dynamical phase
\begin{equation}
\langle \mu,k_{\mathrm{dyn}}(s) \rangle = - \int_{0}^{2\pi} H_{t} (\mathrm{Ad}_{p^{-1}(s,t)}^{\ast}\mu)
\der {t}.       \label{DPh}%
\end{equation}
Here $p(s,t)$ is defined by (\ref{Pe1}). Let $\mathcal{O} \subset \mathfrak{g}^{\ast}$ be the
co-adjoint orbit passing through the point $\mu$. Fix $s \in [0,1]$ and consider the oriented
cylinder $\mathcal{C}_{s}^{2} = [0,s] \times \mathbb{S}^{1}$ with coordinates $(s,u \; \mathrm{mod} \, 2\pi
)$. Define the $C^{\infty}$ mapping $F : \mathcal{C}_{s}^{2} \rightarrow \mathfrak{g}^{\ast}$ by
$F(u,t) = \mathrm{Ad}_{p^{-1}(u,t)}^{\ast} \mu$. It is clear that the image of $\mathcal{C}_{s}^{2}$
under $F$ lies in the co-adjoint orbit, $F(\mathcal{C}_{s}^{2}) \subset \mathcal{O}$. Moreover,
$F(\{0\} \times \mathbb{S}^{1}) = \mu$. Let $\omega_{\mathcal{O}}$ be the symplectic form
(the Kirillov form) on $\mathcal{O}$ which is given by
\begin{equation}
\omega_{\mathcal{O}}(\mathrm{ad}_{x}^{\ast} \eta,\mathrm{ad}_{y}^{\ast} \eta) =
\langle \eta,[x,y] \rangle,        \label{Sym}%
\end{equation}
for $\eta \in \mathcal{O}$ and $x,y \in \mathfrak{g}$. Taking into account properties
(\ref{Pr2}), (\ref{Pr4}), we compute
\[
\frac{\partial F(u,t)}{\partial t} = - \mathrm{ad}_{\mathrm{D}_{t}p(u,t)}^{\ast} F(u,t),
\]%
\[
\frac{\partial F(u,t)}{\partial u} = - \mathrm{ad}_{\mathrm{D}_{u}p(u,t)}^{\ast} F(u,t).
\]
Putting these formulas into (\ref{Sym}) and using again (\ref{Pr2}), we get
\begin{align*}
\omega_{\mathcal{O}} \left( \frac{\partial F}{\partial u},\frac{\partial F}{\partial
t} \right) &  = \langle F(u,t), [\mathrm{D}_{u}p(u,t),\mathrm{D}_{t}p(u,t)] \rangle \\
&  = \langle \mathrm{Ad}_{p^{-1}(u,t)}^{\ast}\mu,[\mathrm{D}_{u}p(u,t),\mathrm{D}_{t}p(u,t)] \rangle \\
&  = \langle \mu,[\mathrm{D}_{u}p^{-1}(u,t),\mathrm{D}_{t}p^{-1}(u,t)] \rangle.
\end{align*}
Comparing this with (\ref{Lo3}) leads to the following representation for the geometric phase%
\begin{equation}
\langle \mu,k_{\mathrm{geom}}(s) \rangle = \int_{\mathcal{C}_{s}^{2}}
F^{\ast}\omega_{\mathcal{O}}.                                         \label{GPh}%
\end{equation}
If the mapping $F$ is regular, then the right hand side of (\ref{GPh}) is the symplectic area of
the oriented surface $\Sigma_{s}=F(\mathcal{C}_{s}^{2})$ in $\mathcal{O}$ whose boundary is the
loop $\gamma_{s} = \{\xi = \mathrm{Ad}_{p^{-1}(u,s)}^{\ast}\mu\}$. Notice that formulas (\ref{DPh})
and (\ref{GPh}) remain also valid if $\mu = \mu(s)$ varies smoothly with $s$ and lies at a
co-adjoint orbit $\mathcal{O}$ for all $s \in [0,1]$. In the case when
$\mathrm{ad}_{k(s)}^{\ast}\mu(s) = 0$, the parametrized curves $\gamma_{s}$ are periodic solutions
of linear Euler system (\ref{CoE}) and the values $\langle \mu(s),k_{\mathrm{dyn}}(s) \rangle$ and
$\langle \mu(s),k_{\mathrm{geom}}(s) \rangle$ correspond to the dynamic and geometric parts in the
splitting of Floquet exponents of cyclic solutions of (\ref{CoE}) (for more details, see \cite{FV,
Vor}).

\begin{remark}
If instead of {\upshape (\ref{Hom1})} we have
\[
\phi(s, t + T(s)) = \phi(s,t)
\]
for a certain smooth positive function $T(s)$, then the geometric phase of the corresponding
family of periodic Lie systems is given by {\upshape (\ref{GPh})} and the formula for the dynamic
phase is modified as follows
\begin{equation}
\langle \mu,k_{\mathrm{dyn}}(s) \rangle =
- \int_{0}^{T(s)}H_{t}(\mathrm{Ad}_{p^{-1}(s,t)}^{\ast}\mu) \der {t}.   \label{DynPhchar}
\end{equation}
Here $p(s,t)=L_{f(s,t)} \exp \left(-\frac{tk(s)}{T(s)} \right)$ is a $T(s)$-periodic in $t$
for each $s$.
\end{remark}

Now, using the above results, we introduce dynamical and geometric phases for an
\textit{individual} periodic Lie system (\ref{Li2}) satisfying reducibility condition (\ref{Re1})
for a certain element $k \in \mathfrak{g}$. Consider the loop $\Gamma : t \mapsto p(t)$ in $G$,
based at $e$, where $p(t) = f(t) \exp(-{tk}/{2\pi})$ is the $2\pi$-periodic, $G$-valued function
corresponding to the periodic factor in the Floquet representation. Then, $\Gamma$ depends on the
choice of $k$ in (\ref{Re1}), but it is easy to see that the homotopy class $[\Gamma]$ of $\Gamma$
in $\pi_{1}(G)$ is independent of any such choice. Assume that, $[\Gamma]$ is \textit{trivial}. In
this case, we say that the reducible periodic Lie system is \textit{contractible}. This condition
means that we can fix a smooth homotopy in $G$ of  the loop $\Gamma$ to the unity $e$ which is
given by a $C^{\infty}$ function $p(s,t)$ satisfying (\ref{Pe4}) and $p(1,t) = p(t)$. Pick an
arbitrary $C^{\infty}$ function $k(s)$ on $[0,1]$ with $k(0) = 0$ and $k(1) = k$. For example, one
can put $k(s) = sk$. Then, we define
\begin{equation}
\phi(s,t) := \frac{1}{2\pi} \mathrm{Ad}_{p(s,t)}k(s) + \mathrm{D}_{t}p(s,t).   \label{Nov}%
\end{equation}
Clearly, this function satisfies properties (\ref{Hom1}), (\ref{Hom2}) and $\phi(1,t) = \phi(t)$.
Therefore, we have proved that the original Lie system is included into a smooth family of reducible
periodic Lie systems on $G$ associated with $\phi$ in (\ref{Nov}) which is contractible to the trivial
system $\dot{g} = 0$. Applying formulas (\ref{DPh}), (\ref{GPh}) to this family, we arrive at the
final result.

\begin{theorem}\label{MainTheorem}
Assume that a periodic Lie system
\[
\frac{\der {g}}{\der {t}} = T_{e}R_{g}(\phi(t)), \qquad g\in G
\]
is reducible, $m \in \exp(\mathfrak{g})$, and contractible. Let $p(s,t)$ be an arbitrary smooth
homotopy of $\Gamma$ to the identity $e$. Then, the monodromy element has the representation
\[
m = \exp(k_{\mathrm{dyn}} + k_{\mathrm{geom}}),
\]
where the dynamic and geometric phases $k_{\mathrm{dyn}}, k_{\mathrm{geom}} \in \mathfrak{g}$
are given by
\begin{align}
\langle \mu, k_{\mathrm{dyn}} \rangle & = - \int_{0}^{2\pi}H_{t}(F(1,t)) \der {t}, \label{Dyn}\\
  &  \nonumber\\
\langle \mu, k_{\mathrm{geom}} \rangle & = \int_{[01]\times\mathbb{S}^{1}}
   F^{\ast}\omega_{\mathcal{O}}
\end{align}
for any $\mu \in \mathfrak{g}^{\ast}$. Here $F(s,t) = \mathrm{Ad}_{p^{-1}(s,t)}^{\ast}\mu$,
$H_{t}(\xi) = - \langle \xi,\phi(s,t) \rangle$ and $\mathcal{O}$ is the co-adjoint orbit through $\mu$.
\end{theorem}

Remark that the dynamic phase (\ref{Dyn}) is independent of the choice of a homotopy $p(s,t)$. The
elements $k_{\mathrm{dyn}}$ and $k_{\mathrm{geom}}$ can be also called the dynamical and geometric
phases of the periodic Lie system (\ref{Li1}) in the $G$-space $(M,G,\Phi)$.

\section{Some Applications}

Consider the following dynamical system in $\mathfrak{g}^{\ast}\times G$%

\begin{eqnarray}
\frac{\der \xi}{\der {t}} &=&   - \mathrm{ad}_{\frac{\delta h}{\delta\xi}}^{\ast}\xi,    \label{EN1}\\
\frac{\der {g}}{\der {t}} &=&   T_{e}R_{g} \left(\frac{\delta h}{\delta\xi} \right),\label{EN2}%
\end{eqnarray}
where $\xi \in \mathfrak{g}^{\ast}$, $g \in G$ and $h : \mathfrak{g}^{\ast} \rightarrow \mathbb{R}$ is
a $C^{\infty}$ function and the element $\frac{\delta h}{\delta\xi} \in\mathfrak{g}$ is defined by the
equality $\langle \mu,\frac{\delta h}{\delta\xi} \rangle = d_{\xi} h(\mu)$ \cite{MR}. This system comes
from a $G$-invariant Hamiltonian system in $T^{\ast}G$ under the identification of
$T^{\ast}G$ with $\mathfrak{g}^{\ast}\times G$ by the right translations \cite{MR}. Suppose we are
given a one-parameter family of periodic trajectories $\gamma_{s} : t \mapsto \xi_{s}(t)$ of nonlinear
Euler system (\ref{EN1}), $\xi_{s}(t + T(s)) = \xi_{s}(t)$ which are smoothly contractible to a rest
point $\xi_{0}(t) = \eta_{0}$, $d_{\eta_{0}}h = 0$. Putting these periodic solutions in second
equation (\ref{EN2}), we get a family of periodic Lie systems on $G$ associated with the
$\mathfrak{g}$-valued function
\begin{equation}
\phi(s,t) = \frac{\delta h(\xi_{s}(t))}{\delta\xi}.                   \label{EN3}%
\end{equation}
which satisfies the conditions (\ref{Hom1}) and (\ref{Hom2}). Let $f(s,t)$ be the fundamental solution of
(\ref{EN2}), (\ref{EN3}) and $m(s) = f(s,T(s))$ the monodromy element. Assume that the periodic
orbits belong to one and the same co-adjoint orbit $\mathcal{O}$,%
\[
\xi_{s}^{0} := \xi_{s}(0) \in\mathcal{O}, \qquad \forall \, s \in [0,1].
\]
Remark that $\xi_{s}(t)$ is also a periodic solution of periodic linear Euler system
(\ref{CoE}) with $\phi$ given by (\ref{EN3}). This implies that for every $s\in [0,1]$, the
monodromy element $m(s)$ lies in the isotropy subgroup of the co-adjoint representation at
$\xi_{s}^{0}$,
\[
m(s) \in G_{\xi_{s}^{0}} = \{\alpha\in G \; | \; \mathrm{Ad}_{\alpha}^{\ast} \xi_{s}^{0} = \xi_{s}^{0} \}.
\]
The Lie algebra of $G_{\xi_{s}^{0}}$ is the isotropy
$\mathfrak{g}_{\xi_{s}^{0}} = \{a \in \mathfrak{g} \; | \;\mathrm{ad}_{a}^{\ast}\xi_{s}^{0} = 0\}$.
Moreover, we assume that $G_{\xi_{s}^{0}}$ is \textit{connected} and the co-adjoint orbit $\mathcal{O}$
is \textit{regular}. Then, $G_{\xi_{s}^{0}}$ is Abelian and exponential \cite{DK}. It follows that
there exists a unique $C^{\infty}$ curve $[0,1] \ni s \mapsto k(s) \in \mathfrak{g}$ such that
$k(s) \in \mathfrak{g}_{\xi_{s}^{0}}$, $k(0) = 0$ and $m(s) = \exp k(s)$. Applying the results
of Section \ref{sec:dyngeom}, we get that the total $\log$ phase $k(s)$ has decomposition (\ref{Lo1}),
where $k_{\mathrm{geom}}(s)$ and $k_{\mathrm{dyn}}(s)$ are given by formulae (\ref{GPh}) and
(\ref{DPh}), respectively. Taking into account that
$\xi_{s}(t) = \mathrm{Ad}_{p^{-1}(s,t)}^{\ast}\xi_{s}^{0}$ and evaluating $k_{\mathrm{geom}}(s)$ at
$\mu = \xi_{s}^{0}$, we get
\begin{align}
\langle \xi_{s}^{0}, k_{\mathrm{dyn}}(s) \rangle & = \int_{0}^{T(s)} \langle \xi,
\frac{\delta h}{\delta\xi} \rangle \big|_{\xi=\xi_{s}(t)} \der {t},                  \label{Rec1}\\
  &  \nonumber\\
\langle \xi_{s}^{0}, k_{\mathrm{geom}(s)} \rangle & = \int_{\Sigma_{s}} \omega_{\mathcal{O}},
   \label{Rec2}%
\end{align}
where $\displaystyle \Sigma_{s} = \bigcup_{0 \leq s^{\prime} \leq s} \gamma_{s^{\prime}}$ is the oriented
surface in $\mathcal{O}$ spanned by the periodic trajectories. In particular, if $h$ is a quadratic
form, then $\langle \xi,\frac{\delta h}{\delta\xi} \rangle = 2h(\xi)$ and%
\begin{equation}
\langle \xi_{s}^{0}, k_{\mathrm{dyn}} \rangle = 2T(s)h(\xi_{s}^{0}).              \label{Rec3}%
\end{equation}
Here, we use the property that $h$ is constant along $\gamma_{s}$. In the context of the theory of
reconstruction phases for simple mechanical systems, formulas like (\ref{Rec2}) and (\ref{Rec3})
were derived in \cite{Bla, MMR}. In the case when $G = {SO}(3)$, these formulas lead to the
well-known
representations for the rigid body phases \cite{M}.

\noindent\textbf{Acknowledgements}. The authors thank G. D\'avila-Rasc\'on for fruitful discussions
and comments. This research was partially supported by CONACYT under the grant no. 55463. JdL acknowledges a FPU grant
from Ministerio de Educacion y Ciencia and partial support by the research projects MTM2006-10531 and E24/1 (DGA).

\noindent Rub\'en Flores Espinoza, rflorese@gauss.mat.uson.mx\\
Javier de Lucas, delucas@impan.gov.pl\\
Yuri M. Vorobiev, yurimv@guaymas.uson.mx

\end{document}